\PassOptionsToPackage{normalem}{ulem}
\PassOptionsToPackage{table,dvipsnames}{xcolor}
\documentclass[]{arxiv_template}

\usepackage[utf8]{inputenc}
\usepackage{times}
\usepackage{latexsym}
\usepackage{inconsolata}
\usepackage{graphicx}
\usepackage{booktabs}
\usepackage{xspace}
\usepackage{amsmath}
\usepackage{amssymb}
\usepackage{subcaption}

\definecolor{promptbg}{HTML}{F4FAF8}
\definecolor{promptframe}{HTML}{B8E0D8}
\definecolor{prompttitle}{HTML}{2EC4B6}

\newtcolorbox{promptbox}[1]{
  enhanced,
  breakable,
  colback=promptbg,
  colframe=promptframe,
  colbacktitle=prompttitle,
  coltitle=white,
  title={\strut\textbf{#1}},
  fonttitle=\sffamily\bfseries\small,
  fontupper=\ttfamily\footnotesize,
  arc=4pt,
  boxrule=0.4pt,
  left=8pt, right=8pt, top=5pt, bottom=5pt,
  toptitle=3pt, bottomtitle=3pt,
  drop shadow={black!8},
}

\newcommand{\approach}{Dockerless\xspace}

\title{Dockerless: Environment-Free Program Verifier for Coding Agents}

\makeatletter
\gdef\authorlist{%
  \begin{center}
    \authorfont
    Wenhao Zeng$^{1,*}$\quad
    Yuling Shi$^{1,*}$\quad
    Xiaodong Gu$^{1,\dagger}$\quad
    Chao Hu$^{1}$\\[2pt]
    Chaofan Wang$^{1}$\quad
    Yuhao Cui$^{2}$\quad
    Hongting Zhou$^{2}$\quad
    Mengnan Qi$^{2}$\\[2pt]
    Jianqiao Wangni$^{2}$\quad
    Zhaojian Yu$^{2}$\quad
    Shuzheng Gao$^{2}$\quad
    Kai Cai$^{2}$\quad
    Shilin He$^{2,\dagger}$
  \end{center}
}
\makeatother

\makeatletter
\gdef\affiliationlist{%
  \begin{center}
    \affiliationfont
    $^{1}$Shanghai Jiao Tong University
    $^{2}$Douyin Group
  \end{center}
}
\makeatother

\abstract{
Program verifiers play a central role in training coding agents, including selecting trajectories for supervised fine-tuning (SFT) and providing rewards for reinforcement learning (RL).
  Standard execution-based verification requires running unit tests inside per-repository environments such as Docker images, incurring substantial environment setup costs.
  We propose \approach, an environment-free agentic patch verifier that evaluates generated code patches without executing them.
  Rather than simply matching candidate patches to references, \approach judges patch correctness using evidence gathered through agentic repository exploration.
  On a verifier evaluation benchmark, \approach outperforms the strongest open-source verifier by $14.3$ AUC points.
  Using \approach as both the SFT trajectory filter and the RL reward enables a fully environment-free post-training pipeline.
  The resulting model reaches $62.0\%$, $50.0\%$, and $35.2\%$ resolve rate on SWE-bench Verified, Multilingual, and Pro, respectively.
  It surpasses the Qwen3.5-9B baseline by $2.4$, $8.7$, and $2.9$ points, matching environment-based post-training.

}

\begin{document}
\maketitle

{
\renewcommand{\thefootnote}{\fnsymbol{footnote}}
  \footnotetext[1]{Equal contribution.}
  \footnotetext[2]{Corresponding authors: \email{xiaodong.gu@sjtu.edu.cn} and \email{heshilin@bytedance.com}}
}

\section{Introduction}
\label{sec:intro}

Program verifiers play a critical role in training automated coding agents. Whether curating high-quality trajectories for supervised fine-tuning (SFT)~\citep{yuan2023scaling,pan2025training,jain2025r2e,zeng2025pruning} or providing rewards for reinforcement learning (RL)~\citep{wei2025swerl,luo2025deepswe}, verifiers determine whether the agent rollouts successfully resolve issues. Currently, the gold standard for this correctness feedback relies on executing test cases inside isolated, per-repository environments~\citep{jimenez2024swe,pan2025training,jain2025r2e}.

However, execution-based verification imposes substantial engineering overhead.
Setting up these environments requires building custom Docker images, resolving per-repository dependencies, identifying relevant tests, and writing test-execution scripts and result parsers.
Even advanced automated pipelines still succeed on only a limited share of candidate repositories~\citep{jain2025r2e,badertdinov2026swe,li2026repolaunch,zhang2025swebenchgoeslive}.
More fundamentally, many real-world repositories, especially private, enterprise, or legacy codebases, lack reproducible environments or comprehensive test suites altogether~\citep{pan2025training,zan2026multi}, making execution-based verification unreliable or infeasible.

\begin{figure}[t]
  \centering
  \includegraphics[width=0.70\linewidth]{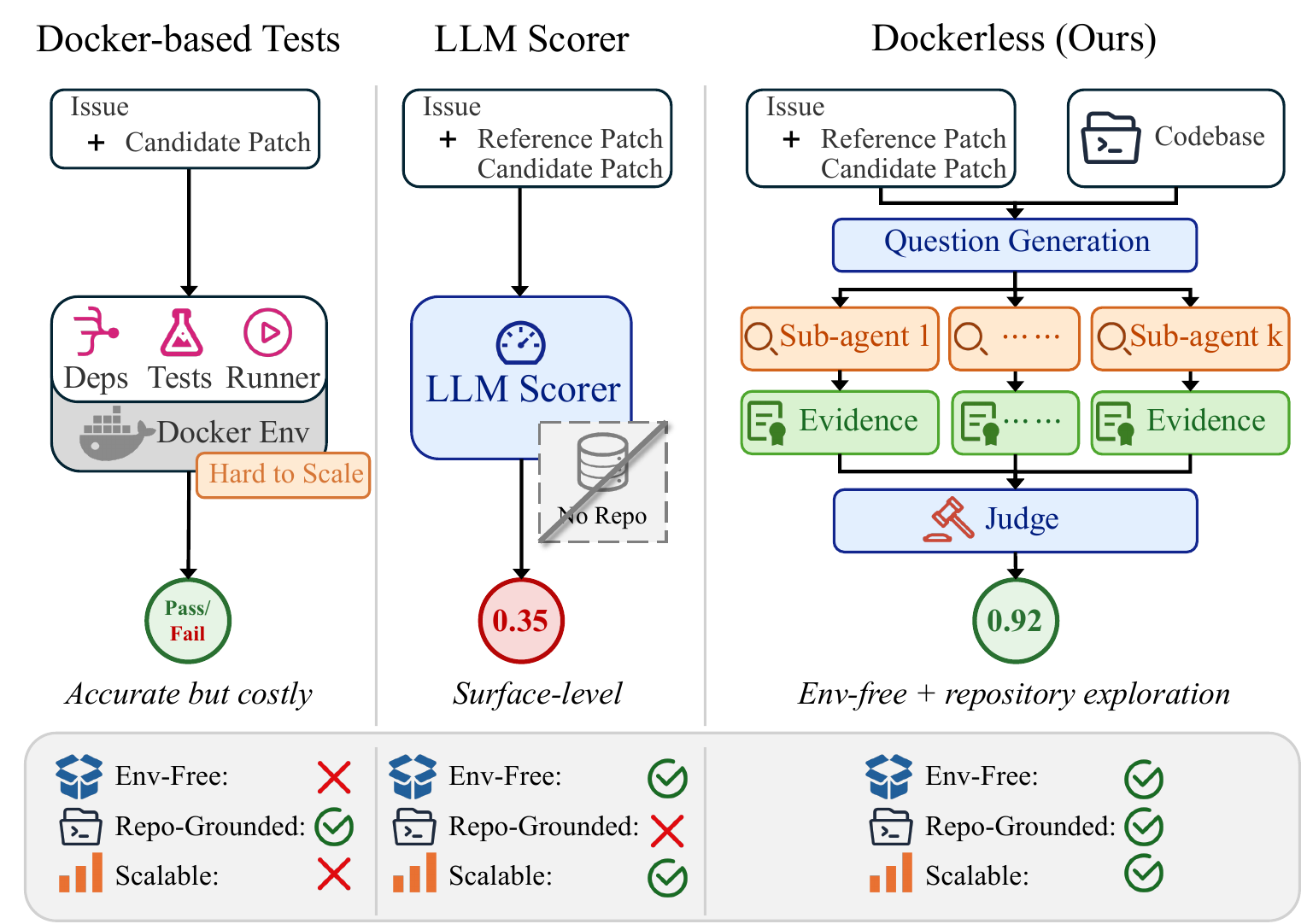}
  \caption{Comparison of verifiers for SWE agents. Docker-based tests are accurate but depend on costly per-repository environments. LLM scorers sidestep that cost but score patches based on surface-level information, without actively inspecting the repository. \approach instead deeply explores the codebase to judge the patch, requiring no per-repository environment while retaining repository grounding.}
  \label{fig:comparison}
\end{figure}

To reduce setup costs, recent work executes agent rollouts from a single shared base image rather than per-repository Docker containers~\citep{sun2026swe,ludwig2026swe,xu2025scalable}.
Yet, the verifier remains a critical bottleneck.
Existing environment-free verifiers score patches using only surface-level information, without ever inspecting the repository~\citep{shum2025swe,wang2026rubric,luo2025deepswe}.
Such shallow approaches are insufficient for complex SWE tasks, where determining functional equivalence requires deep repository context: for example, whether a modified function is actually called by the failing behavior, or whether an alternative implementation correctly integrates with surrounding modules.

To close this gap, we propose \approach, an environment-free agentic verifier that actively explores the repository to judge patch correctness. As shown in \cref{fig:comparison}, rather than blindly matching textual diffs, \approach grounds its verification in the actual codebase. 
Given an issue description, a reference patch, and a candidate patch, \approach first derives several verification questions from the issue and the reference patch. It then dispatches dedicated sub-agents to gather repository evidence for each question.
Finally, it aggregates the collected evidence into a correctness score indicating whether the candidate patch correctly resolves the issue.
We train \approach via rejection sampling on $3.7$K issues from SWE-Gym~\citep{pan2025training} and Multi-SWE-RL~\citep{zan2026multi}, retaining only question-answer-judge trajectories whose final verdict matches the ground-truth test outcome.

Ultimately, \approach unlocks a fully environment-free post-training pipeline: rollout collection, SFT data filtering, and RL reward computation can all run on a minimal base image with \textbf{zero per-repository setup}.
As a standalone verifier, \approach outperforms the strongest open-source baseline by $14.3$ AUC points on a verifier evaluation benchmark.
For SFT, training on the top $25\%$ of trajectories filtered by \approach ($4$K out of $16$K) surpasses training on the full environment-free pool by $1.8$, $6.4$, and $3.4$ points on SWE-bench Verified, Multilingual, and Pro, respectively.
For RL, using \approach as an environment-free reward outperforms RL with the DeepSWE Verifier by $1.4$, $2.7$, and $1.1$ points on the same three benchmarks.
End-to-end, our fully environment-free post-training pipeline produces a model that reaches $62.0\%$, $50.0\%$, and $35.2\%$ resolve rate on SWE-bench Verified, Multilingual, and Pro~\citep{jimenez2024swe,openai2024introducing,yang2025swe,deng2025swe}, improving over the Qwen3.5-9B baseline by $2.4$, $8.7$, and $2.9$ points, respectively.
By matching the performance of standard environment-based post-training, \approach establishes environment-free post-training as a scalable and viable path for the vast long tail of real-world repositories.

\begin{itemize}
  \item We propose \approach, an environment-free agentic verifier that scores patches by actively exploring the repository with parallel sub-agents.
  \item By providing reliable correctness feedback, \approach enables a fully environment-free post-training pipeline for SFT trajectory filtering and RL rewards, scaling coding-agent post-training.
  \item Empirically, \approach outperforms the strongest open-source verifier by $14.3$ AUC points, while the resulting fully environment-free post-training pipeline achieves performance comparable to standard environment-based post-training.
\end{itemize}

\begin{figure}[th]
  \centering
  \includegraphics[width=0.95\linewidth]{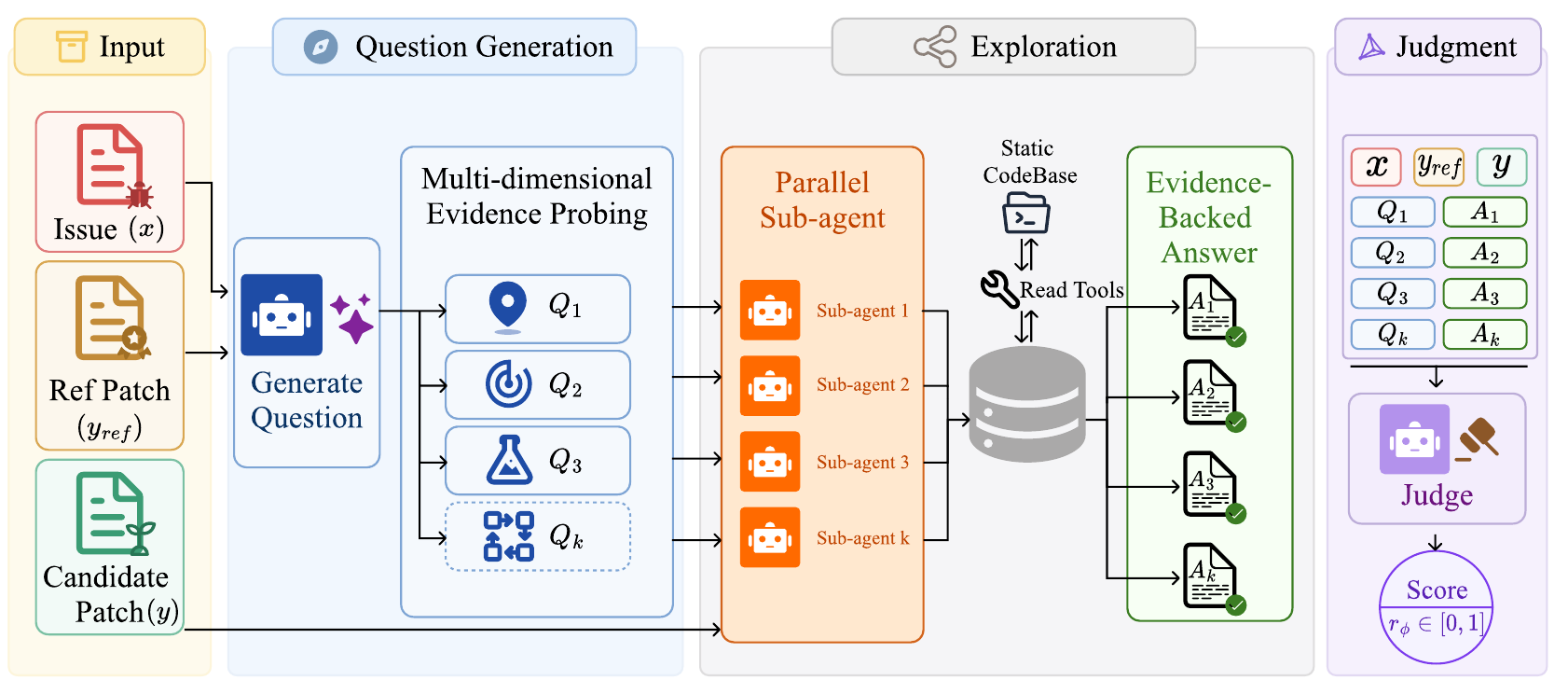}
  \caption{Architecture of \approach. The verifier takes the issue $x$, reference patch $y_{\text{ref}}$, and candidate patch $y$, and proceeds in two stages.
    (1) Question generation and exploration: the verifier first generates $K$ verification questions and dispatches parallel sub-agents to collect evidence-backed answers from the codebase.
    (2) Judgment: the verifier conditions on the issue, the patches, and the collected $(Q_k, A_k)$ pairs to produce a binary verdict token, whose logits define the continuous score $r_\phi(x, y)$.}
  \label{fig:method}
\end{figure}


\section{Methodology}
\label{sec:method}

\subsection{Problem Setting}
\label{sec:motivation-setting}
Given an issue $x$ and a candidate patch $y$, a verifier assigns a correctness score $r(x, y) \in [0, 1]$ indicating whether \(y\) resolves \(x\).

In standard SWE post-training, which we call the \emph{environment-based} (\emph{env-based}) setting, candidate patches are verified by executing held-out tests inside a repository-specific environment ($E_x$). $E_x$ consists of a Docker image with pinned dependencies, a curated unit-test suite, and a working test runner.
This produces a binary correctness signal:
\begin{equation}
  r_{\text{env}}(x, y) = \mathbb{1}\!\left[\,\text{tests in } E_x \text{ pass under } y\,\right].
  \label{eq:env-reward}
\end{equation}
However, building these environments is prohibitively expensive, and many real-world codebases lack reproducible environments or usable test suites.


To make post-training scalable, we consider the \emph{environment-free} (\emph{env-free}) setting in which agents run in a single minimal base image without repository-specific dependencies, test runners, or access to \(E_x\).
This setting is already practical on the agent side: frontier models under the OpenHands scaffold retain much of their performance after removing the per-repository environment, with resolve-rate drops of $3.0$--$13.9$ points (\cref{app:motivation}).
Thus, env-free rollouts can be collected at scale; the remaining bottleneck is verification.
Our goal is to train an environment-free verifier \(r_\phi(x,y)\) that can replace \(r_{\text{env}}\) for both SFT trajectory filtering and RL reward computation.

\subsection{Architecture of \approach}
\label{sec:agentic-verifier}

As illustrated in \cref{fig:method}, the verifier operates in two stages.
First, given an issue \(x\) and a reference patch \(y_{\text{ref}}\), the model proposes a small set of verification questions $\{Q_1, \dots, Q_K\}$. These questions ask, for example, \textit{where in the repository the fix should take effect}, \textit{what the patched code is supposed to do}, \textit{ what tests or assertions would confirm correctness}, and \textit{whether other parts of the repository could break}.
Answering these questions grounds the verifier's eventual judgment in repository exploration rather than in surface-level comparison between the candidate and the reference patch.
For each question, a sub-agent then explores the repository through read-only shell tools (e.g., find, grep, rg) and returns a short evidence-backed answer $A_k$.
The $K$ sub-agents run in parallel for efficiency.

After all sub-agents return their answers, \approach aggregates the collected evidence to judge whether the candidate patch \(y\) resolves the issue \(x\).
Given \((x, y_{\text{ref}}, y, \{(Q_k, A_k)\}_{k=1}^{K})\), the verdict model outputs a binary token in \(\{0,1\}\), where \(1\) denotes a correct patch.
At inference time, we convert the logits of the two verdict tokens into a continuous score:
\[
  r_\phi(x, y) =
  \frac{\exp(\ell_1)}{\exp(\ell_0) + \exp(\ell_1)},
\]
where \(\ell_0\) and \(\ell_1\) are the logits for tokens \(0\) and \(1\).
The full prompts used at both stages are listed in~\cref{app:prompts}.


\begin{figure}[t]
  \centering
  \includegraphics[width=0.60\linewidth]{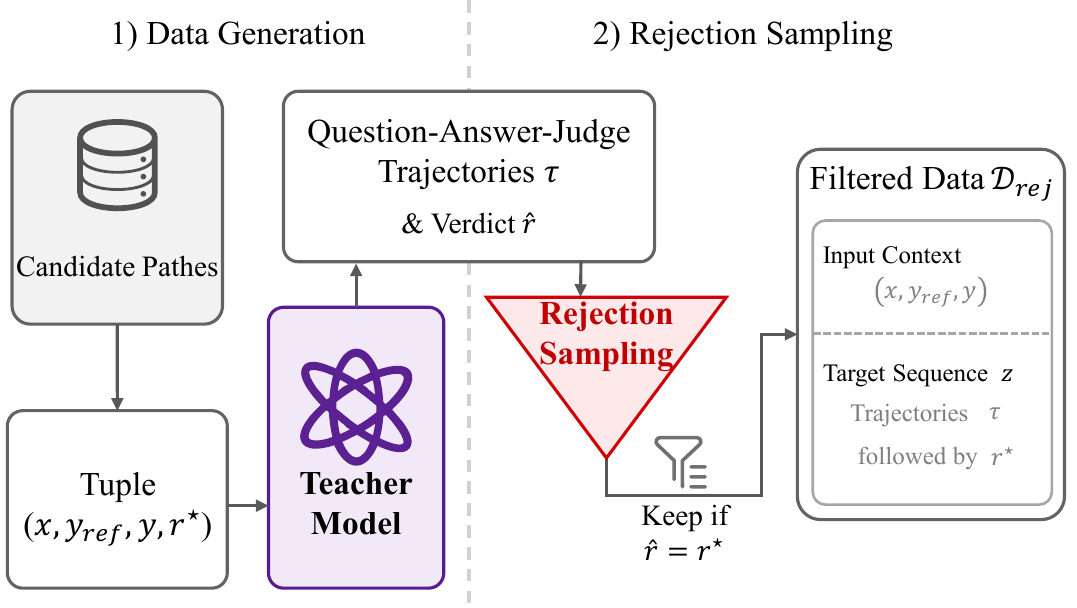}
  \caption{Training pipeline for \approach: teacher-generated question-answer-judge trajectories are rejection-sampled by matching the predicted verdict against the ground-truth, and used to fine-tune a base model.}
  \label{fig:SFT}
\end{figure}

\subsection{\approach Training}
\label{sec:verifier-training}

We train the verifier $r_\phi$ via rejection sampling on execution-labeled candidate patches.
Each example is a tuple $(x, y_{\text{ref}}, y, r^\star)$, where $r^\star \in \{0, 1\}$ is the ground-truth verdict obtained by running the held-out unit tests on the candidate patch $y$.
\cref{fig:SFT} gives an overview of the training pipeline.

To construct training trajectories $\tau$ for the question-answer-judge, an agent powered with a teacher model explores the repository until making a judgment verdict \(\hat{r}\in\{0,1\}\) for each example.
We then reject-sample these trajectories, keeping only those whose $\hat{r}$ matches the execution label $r^\star$; the retained examples form \(\mathcal{D}_{\text{rej}}\).
This keeps the training signal consistent end-to-end, and the verifier learns how to reason step-by-step and conclude the final verdict rather than from lucky matches.
We additionally cap the negative-to-positive sample ratio at $\rho$ to mitigate class imbalance, following the recipe of \citet{shum2025swe}.

The verifier is then trained with the standard next-token cross-entropy over the full output sequence.
\begin{equation}
  \mathcal{L}_\phi
  =
  -\mathbb{E}_{\mathcal{D}_{\text{rej}}}
  \left[
    \sum_{t=1}^{T}
    \log p_\phi(z_t \mid x, y_{\text{ref}}, y, z_{<t})
    \right],
  \label{eq:rm-loss}
\end{equation}
where \(z=(z_1,\ldots,z_T)\) denotes the token sequence in the question-answer-judge trajectories \(\tau\).
A single backbone is shared across question generation, sub-agent exploration, and the final judging stage, jointly trained under \cref{eq:rm-loss}.
Full training details are in \cref{app:training-verifier}.

\subsection{Environment-Free Post-training}
\label{sec:post-training}

\begin{figure}[t]
  \centering
  \includegraphics[width=0.70\linewidth]{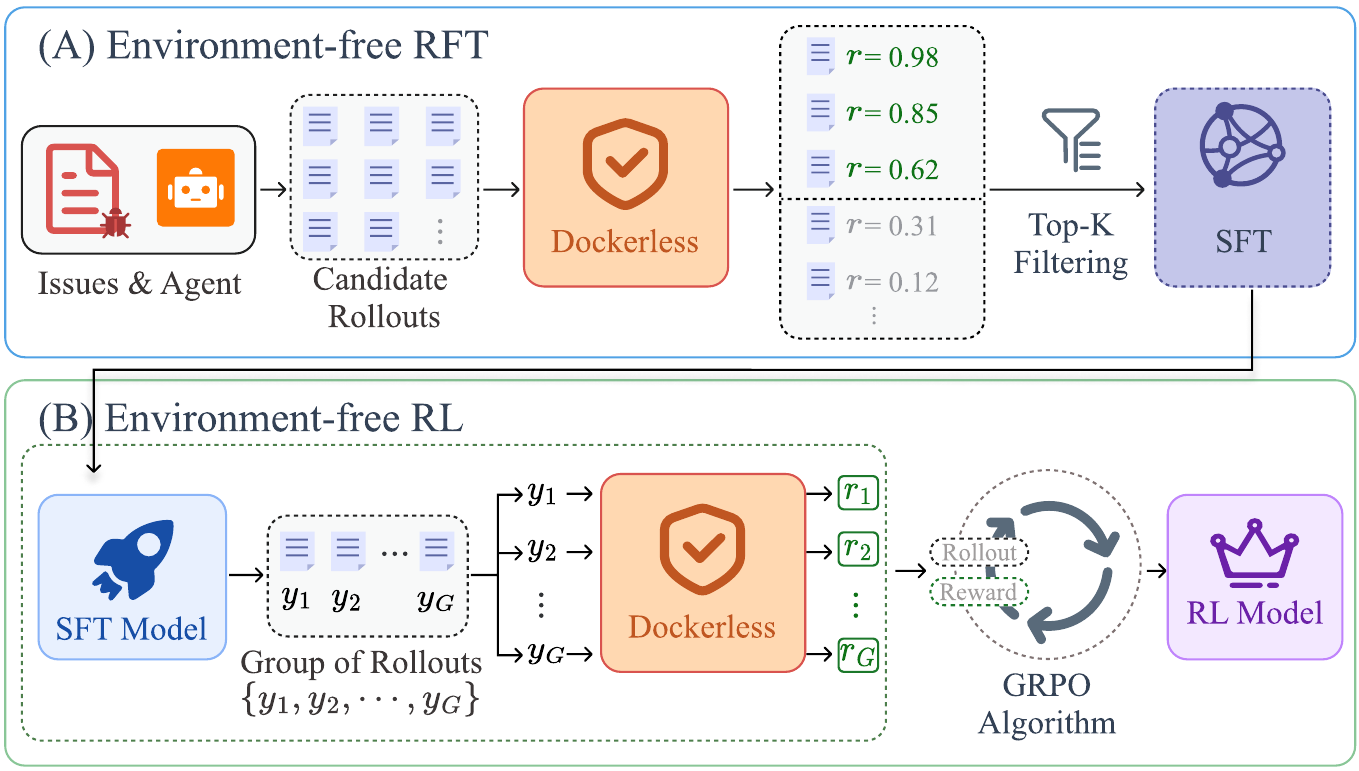}
  \caption{Env-free post-training pipeline for \approach. (A) Environment-free RFT: candidate rollouts are scored by \approach, and the top-$K$ are kept to fine-tune the base model, yielding the SFT model. (B) Environment-free RL: starting from the SFT model, GRPO uses \approach as the per-rollout reward source, yielding the RL model.}
  \label{fig:pipeline}
\end{figure}

With \approach trained, we now apply it in the environment-free post-training pipelines, illustrated in \cref{fig:pipeline}.

\paragraph{Environment-free RFT.}
Rejection-sampling fine-tuning (RFT)~\citep{yuan2023scaling} curates SFT data by keeping only the high-quality rollouts whose final patches pass per-repository unit tests~\citep{pan2025training,jain2025r2e}.
We instead start from an agent, collect a large pool of rollouts in a minimal Linux image without instantiating per-repository environments, and use \approach as the rejection signal.
We score each rollout's final patch with \approach and form $\mathcal{D}_{\text{RFT}}$ by keeping the top-$K$ rollouts globally ranked by $r_\phi$.
We then fine-tune the model on $\mathcal{D}_{\text{RFT}}$ with the standard SFT objective, yielding the SFT model.


\paragraph{Environment-free RL.}
We further use \approach as the reward model for RL on top of the SFT model. During RL, rollouts are collected in the same minimal Linux image used for env-free RFT, i.e., without a per-repository environment.
For each rollout on issue \(x\), let \(y_i\) denote its final patch. We score \(y_i\) with \approach and use \(r_\phi(x,y_i)\) as the reward.
We then optimize the model with GRPO~\citep{shao2024deepseekmath}. For each group of $G$ rollouts on issue $x$, let \(\{y_1,\dots,y_G\}\) denote their final patches. We form group-normalized advantages
\begin{equation}
  A_i = \frac{r_\phi(x, y_i) - \bar{r}}{\hat{\sigma}_r}, \qquad \bar{r} = \frac{1}{G}\sum_{j=1}^G r_\phi(x, y_j),
  \label{eq:grpo-advantage}
\end{equation}
where \(\hat{\sigma}_r\) is the standard deviation of the verifier rewards in the group. These advantages are then used in the standard GRPO objective.
To improve reward stability, we compute each reward by averaging \(M\) independent \approach evaluations of the same final patch.

\section{Experimental Settings}
\label{sec:setup}

\paragraph{Benchmarks.}
For agent resolve rate, we evaluate on SWE-bench Verified~\citep{jimenez2024swe,openai2024introducing}, SWE-bench Multilingual~\citep{yang2025swe}, and SWE-bench Pro~\citep{deng2025swe}.
For evaluating the verifier itself, we follow recent practice~\citep{shum2025swe} and construct a balanced trajectory-level verifier evaluation benchmark of $776$ samples ($500$ from SWE-bench Verified and $276$ from Multi-SWE-bench Flash~\citep{zan2026multi}); construction details are in \cref{app:rm-bench}.
\begin{table}[t]
\centering
\small
\resizebox{0.96\linewidth}{!}{%
\begin{tabular}{l l l c ccc}
\toprule
\textbf{Model}
& \textbf{Base}
& \textbf{Training}
& \textbf{Env-free}
& \textbf{Verified}
& \textbf{Multilingual}
& \textbf{Pro} \\
\midrule
\multicolumn{7}{l}{\textit{Open-source models}} \\
\texttt{SWE-Gym-7B}             & \texttt{Qwen2.5-Coder-7B}            & SFT    & No               & 11.4 & 3.0  & 3.3  \\
\texttt{SWE-Dev-7B}             & \texttt{Qwen2.5-Coder-7B}            & SFT    & No               & 10.6 & 4.2  & 4.0  \\
\texttt{SWE-Lego-8B}            & \texttt{Qwen3-8B}                    & SFT    & No               & 41.2 & 19.0 & 16.1 \\
\texttt{Qwen3.5-9B}             & --                                   & --     & --             & 59.6 & 41.3 & 32.3 \\
\midrule
\multicolumn{7}{l}{\textit{Ours: SFT on varying rollout sources}} \\
\textcolor{gray}{\texttt{Env-SFT-9B}}             & \textcolor{gray}{\texttt{Qwen3.5-9B}}                  & \textcolor{gray}{SFT}    & \textcolor{gray}{No}               & \textcolor{gray}{60.0} & \textcolor{gray}{48.3} & \textcolor{gray}{33.9} \\
\rowcolor{black!4}
\textbf{\texttt{\approach-SFT-9B}} & \texttt{Qwen3.5-9B}               & SFT    & Yes     & \textbf{60.6} & \textbf{47.7} & \textbf{35.3} \\
\midrule
\multicolumn{7}{l}{\textit{Ours: RL on varying rollout and reward sources}} \\
\textcolor{gray}{\texttt{+\,DeepSWE-Verifier RL}}
                                & \textcolor{gray}{\texttt{\approach-SFT-9B}}            & \textcolor{gray}{RL}     & \textcolor{gray}{Yes} & \textcolor{gray}{60.6} & \textcolor{gray}{47.3} & \textcolor{gray}{34.1} \\
\textcolor{gray}{\texttt{+\,Test-Execution RL}}
                                & \textcolor{gray}{\texttt{\approach-SFT-9B}}            & \textcolor{gray}{RL}     & \textcolor{gray}{No}               & \textcolor{gray}{62.4} & \textcolor{gray}{51.3} & \textcolor{gray}{35.7} \\
\rowcolor{black!4}
\textbf{\texttt{\approach-RL-9B}}  & \texttt{\approach-SFT-9B}         & RL     & Yes     & \textbf{62.0} & \textbf{50.0} & \textbf{35.2} \\
\bottomrule
\end{tabular}%
}
\\[2pt]
\caption{
Resolve rate (\%) on SWE-bench Verified, Multilingual, and Pro under env-based evaluation.
\emph{Base} is the starting model of each training stage; 
\emph{Training} marks SFT or RL; 
\emph{Env-free} indicates whether the full training stage avoids per-repository Docker: ``Yes'' uses only a minimal base image, while ``No'' uses per-repository Docker;
Bold rows are our headline models; gray rows are controlled comparisons that isolate the SFT rollout source or the RL rollout and reward source.
}
\label{tab:main}
\end{table}

\paragraph{Evaluation protocol.}
We use OpenHands~\citep{wang2025openhands} as the default agent scaffold with a maximum of $150$ turns.
For env-based evaluation, the agent runs inside the original per-repository Docker image with repository dependencies and test runners.
For env-free evaluation, the agent runs in a minimal Ubuntu 22.04 LTS image (\texttt{ubuntu:jammy-20260109}) with only the repository checkout at the base commit.
The main paper reports env-based evaluation, following the standard SWE evaluation protocol; env-free numbers are deferred to \cref{app:wo-env-eval}.
We report resolve rate for issue resolution.
For verifier evaluation, we follow~\citep{shum2025swe} and report AUC, a discrimination metric aligned with SFT filtering and RL rewards.

\paragraph{Baselines.}
For agent performance, we compare our SFT and RL models against open-source SWE specialists at the same scale (under $10$B parameters): SWE-Gym-7B~\citep{pan2025training}, SWE-Dev-7B~\citep{wang2025swedev}, SWE-Lego-8B~\citep{tao2026swelego}, and the base model Qwen3.5-9B~\citep{team2026qwen35}.
For verifier evaluation, we compare \approach against four frontier LLMs used zero-shot as judges (DeepSeek-V3.2~\citep{liu2025deepseek}, Kimi-K2.5~\citep{team2026kimi}, GLM-5~\citep{zeng2026glm}, GPT-5.4~\citep{openai2026gpt54}) and four trained verifiers: SWE-Gym Verifier~\citep{pan2025training}, R2E-Gym Verifier~\citep{jain2025r2e}, OpenHands Critic~\citep{wang2026rubric}, and DeepSWE Verifier~\citep{luo2025deepswe}.

\paragraph{Implementation details.}
We use Qwen3.5-9B~\citep{team2026qwen35} as the backbone for both \approach and the downstream post-training.
\approach is trained on rejection-sampled trajectories from $3.7$K execution-labeled issues drawn from SWE-Gym~\citep{pan2025training} and Multi-SWE-RL~\citep{zan2026multi}, and uses $K{=}2-4$ verification questions.
For downstream post-training, we use SWE-Rebench-v2~\citep{badertdinov2026swev2}. We collect env-free trajectories for SFT and sample RL rollouts from the same task pool.
Full training data construction and hyperparameters are deferred to \cref{app:data,app:training}.

\section{Results}

\subsection{Main results}
\label{sec:main-results}

\paragraph{Fully environment-free post-training reaches strongest open-source performance.}
Starting from Qwen3.5-9B, our fully environment-free post-training pipeline produces \approach-RL-9B, which reaches $62.0$, $50.0$, and $35.2$ resolve rate on SWE-bench Verified, Multilingual, and Pro, respectively (\cref{tab:main}).
This improves over the base model by $+2.4$, $+8.7$, and $+2.9$ points and over the next-best open-source SWE specialist (SWE-Lego-8B) by $+20.8$, $+31.0$, and $+19.1$ points.

\paragraph{Env-free SFT matches env-based SFT.}
We next isolate the SFT stage by comparing two Qwen3.5-9B SFT models that differ only in the source of their training rollouts: Env-SFT-9B uses trajectories collected with a per-repository environment, while \approach-SFT-9B uses env-free trajectories filtered by \approach.
Despite removing test execution from SFT data filtering, \approach-SFT-9B achieves comparable performance to the env-based baseline ($60.6$ vs.\ $60.0$ on Verified, $47.7$ vs.\ $48.3$ on Multilingual, and $35.3$ vs.\ $33.9$ on Pro).

\paragraph{Env-free RL approaches env-based RL.}
We then isolate the RL stage on top of the same SFT initialization, \approach-SFT-9B.
The three RL variants differ in their rollout environment and reward source: \approach-RL-9B uses env-free rollouts with \approach rewards, DeepSWE-Verifier RL uses DeepSWE Verifier rewards, and Test-Execution RL uses per-repository Docker with oracle test-execution rewards.
\approach-RL-9B achieves performance close to Test-Execution RL ($62.0$ vs.\ $62.4$ on Verified, $50.0$ vs.\ $51.3$ on Multilingual, and $35.2$ vs.\ $35.7$ on Pro), while outperforming DeepSWE-Verifier RL by $+1.4$, $+2.7$, and $+1.1$ points.

\subsection{Verifier evaluation}
\label{sec:verifier-eval}
\begin{table}[t]
\centering
\small
\begin{tabular}{lcc}
\toprule
\textbf{Model}
& \textbf{Verified}
& \textbf{Multi-SWE} \\
\midrule
DeepSeek-V3.2          & 69.4 & 58.5 \\
Kimi-K2.5              & 70.7 & 63.9 \\
GLM-5                  & 73.2 & 62.5 \\
GPT-5.4                & 75.9 & 59.5 \\
\midrule
SWE-Gym Verifier       & 61.0 & 53.7 \\
R2E-Gym Verifier       & 64.3 & 55.1 \\
OpenHands Critic       & 48.6 & 52.2 \\
DeepSWE Verifier       & 66.7 & 62.9 \\
\midrule
\rowcolor{black!4}
\approach              & \textbf{81.0} & \textbf{72.1} \\
\bottomrule
\end{tabular}
\caption{
Verifier AUC on the trajectory-level verifier evaluation benchmark, with splits from SWE-bench Verified and Multi-SWE-bench Flash.
}
\label{tab:rm-bench}
\end{table}

We compare \approach against two baseline families on the balanced trajectory-level verifier evaluation benchmark (\cref{tab:rm-bench}): four frontier LLMs (DeepSeek-V3.2, Kimi-K2.5, GLM-5, GPT-5.4) used zero-shot as judges (\cref{app:prompt-judge}), and four trained open-source verifiers (SWE-Gym Verifier, R2E-Gym Verifier, OpenHands Critic, DeepSWE Verifier).

With agentic repository exploration, \approach reaches $81.0$ AUC on SWE-bench Verified and $72.1$ AUC on Multi-SWE-bench Flash, outperforming every baseline in both splits.
Compared with the strongest trained open-source verifier, \approach improves AUC by $14.3$ points on Verified and $9.2$ points on Multi-SWE-bench Flash; compared with the strongest frontier LLM judge, it improves by $5.1$ and $8.2$ points, respectively.
These results show that the design of \approach, which combines agentic repository exploration with rejection-sampled trajectory training, yields a stronger patch verifier.
This strong verifier performance is the key signal used in \cref{sec:main-results}: \approach filters env-free SFT trajectories and supplies rewards for env-free RL, enabling post-training without per-repository test execution.

\subsection{Effect of the SFT data filter}
\label{sec:sft-filter}
\begin{table}[t]
\centering
\small
\begin{tabular}{lccc}
\toprule
\textbf{Training Data}
& \textbf{Verified}
& \textbf{Multilingual}
& \textbf{Pro} \\
\midrule
None (base) & 59.6 & 41.3 & 32.3 \\
\texttt{All 16K}         & 58.8 & 41.3 & 31.9 \\
\midrule
\texttt{Random 4K}       & 58.2 & 44.3 & 32.0 \\
\texttt{Env-based 4K}    & 60.0 & \textbf{48.3} & 33.9 \\
\rowcolor{black!4} \texttt{\approach 4K} & \textbf{60.6} & 47.7 & \textbf{35.3} \\
\bottomrule
\end{tabular}%
\caption{
Effect of the SFT data filter on downstream resolve rate (\%) under env-based evaluation.
All SFT rows use the same Qwen3.5-9B backbone and SFT recipe; only the selected training data differs.
The base row reports Qwen3.5-9B.
}
\label{tab:sft-same-scale}
\end{table}

We hold the SFT backbone and recipe fixed, and vary only the selected training data (\cref{tab:sft-same-scale}).
Starting from a pool of $16$K env-free trajectories, \texttt{All 16K} trains on the full unfiltered pool, \texttt{Random 4K} samples $4$K trajectories uniformly, and \texttt{\approach 4K} keeps the top-ranked $4$K trajectories selected by \approach.
As an env-based comparison, \texttt{Env-based 4K} uses $4$K trajectories obtained with per-repository environments.

\paragraph{\approach achieves effective trajectory filtering.}
Training on all env-free trajectories does not improve over the base model: \texttt{All 16K} reaches $58.8$, $41.3$, and $31.9$, below or equal to the base on all three benchmarks.
This shows that raw env-free rollouts cannot be used directly for SFT; low-quality trajectories need to be filtered.
\texttt{\approach 4K} substantially outperforms \texttt{Random 4K} on all three benchmarks ($60.6$ vs.\ $58.2$ on Verified, $47.7$ vs.\ $44.3$ on Multilingual, and $35.3$ vs.\ $32.0$ on Pro), 
demonstrating that \approach provides a more effective selection signal than random sampling.

\paragraph{Env-free RFT matches env-based trajectory collection.}
More importantly, env-free trajectory collection combined with \approach filtering achieves performance comparable to SFT on env-based trajectories.
\texttt{\approach 4K} matches \texttt{Env-based 4K} across the three benchmarks ($60.6$ vs.\ $60.0$ on Verified, $47.7$ vs.\ $48.3$ on Multilingual, and $35.3$ vs.\ $33.9$ on Pro).
This suggests a scalable path for RFT: collect rollouts without per-repository setup, then use a strong verifier to select the trajectories worth training on.

\subsection{Effect of the number of verification questions}
\label{sec:question-ablation}

\begin{figure}[t]
\centering
\includegraphics[width=0.68\linewidth]{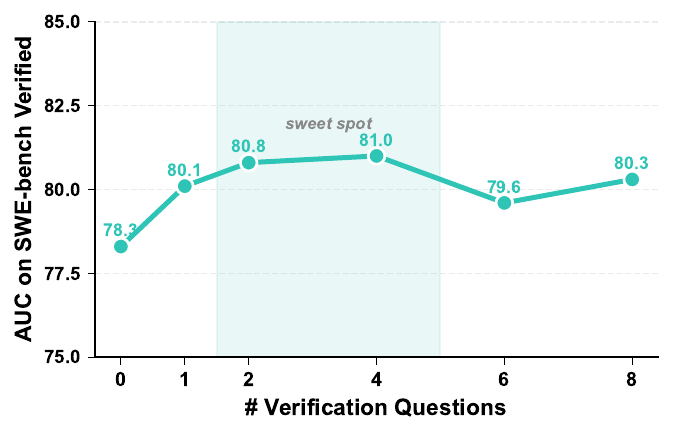}
\caption{
Verifier AUC vs.\ number of verification questions \(K\) on SWE-bench Verified verifier evaluation benchmark.
}
\label{fig:rm-question-ablation}
\end{figure}

\cref{fig:rm-question-ablation} studies how the number of verification questions affects \approach performance.
We vary the number of verification questions \(K \in \{0,1,2,4,6,8\}\) on the SWE-bench Verified split of our verifier evaluation benchmark.
For each setting, \approach first derives \(K\) verification questions from the issue and reference patch, dispatches one sub-agent per question to gather repository evidence, and then judges the candidate patch from the collected Q\&A evidence.
We report AUC against the execution-based ground truth.

\approach performance improves as \(K\) increases from \(0\) to \(4\), rising from $78.3$ AUC with no verification question to $81.0$ AUC at \(K{=}4\).
This shows that asking verification questions and gathering repository evidence helps \approach judge patch correctness.
Beyond four questions, performance fluctuates rather than improving monotonically ($79.6$ at \(K{=}6\), $80.3$ at \(K{=}8\)), suggesting that additional questions often introduce redundant or noisy evidence.
We therefore let \approach generate \(2\)--\(4\) verification questions at inference time, balancing verifier accuracy and per-call exploration cost.



\subsection{Latency analysis}
\label{sec:latency-analysis}

\approach performs a multi-step repository exploration before issuing a reward, so its reward computation is expected to take longer.
We therefore analyze RL training latency under the three reward sources in \cref{sec:main-results}: \approach, the DeepSWE Verifier, and \texttt{Test-Execution}, using $7680$ rollouts.

\begin{figure}[t]
  \centering
  \includegraphics[width=0.70\linewidth]{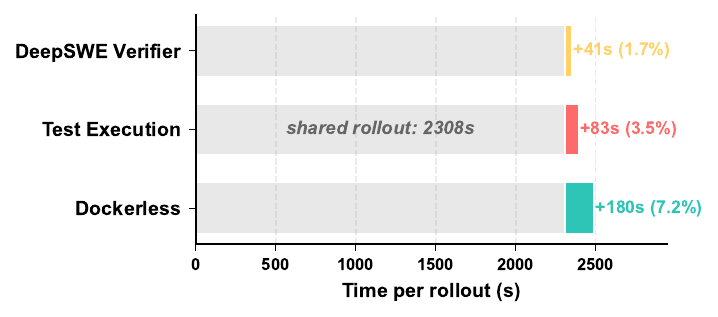}
  \caption{Per-rollout wall-clock breakdown during RL under three reward sources. 
  }
  \label{fig:latency-overhead}
\end{figure}

\cref{fig:latency-overhead} decomposes each RL step into agent rollout time and reward-evaluation time.
Agent rollouts dominate the wall-clock cost, taking $2308$s on average, whereas reward evaluation adds only $41$--$180$s.
Although \approach requires more reward-evaluation time than the other verifier rewards, it still accounts for only $7.2\%$ of the total per-rollout time.
Thus, in the RL setting, the additional cost of agentic verification is small compared with the cost of generating rollouts.
The end-to-end latency distribution shows the same pattern.
As shown in \cref{app:latency}, total per-rollout times under the three reward sources almost overlap, because throughput is dominated by slow rollouts approaching the timeout rather than by reward computation.


\begin{figure}[t]
  \centering
  \includegraphics[width=0.95\linewidth]{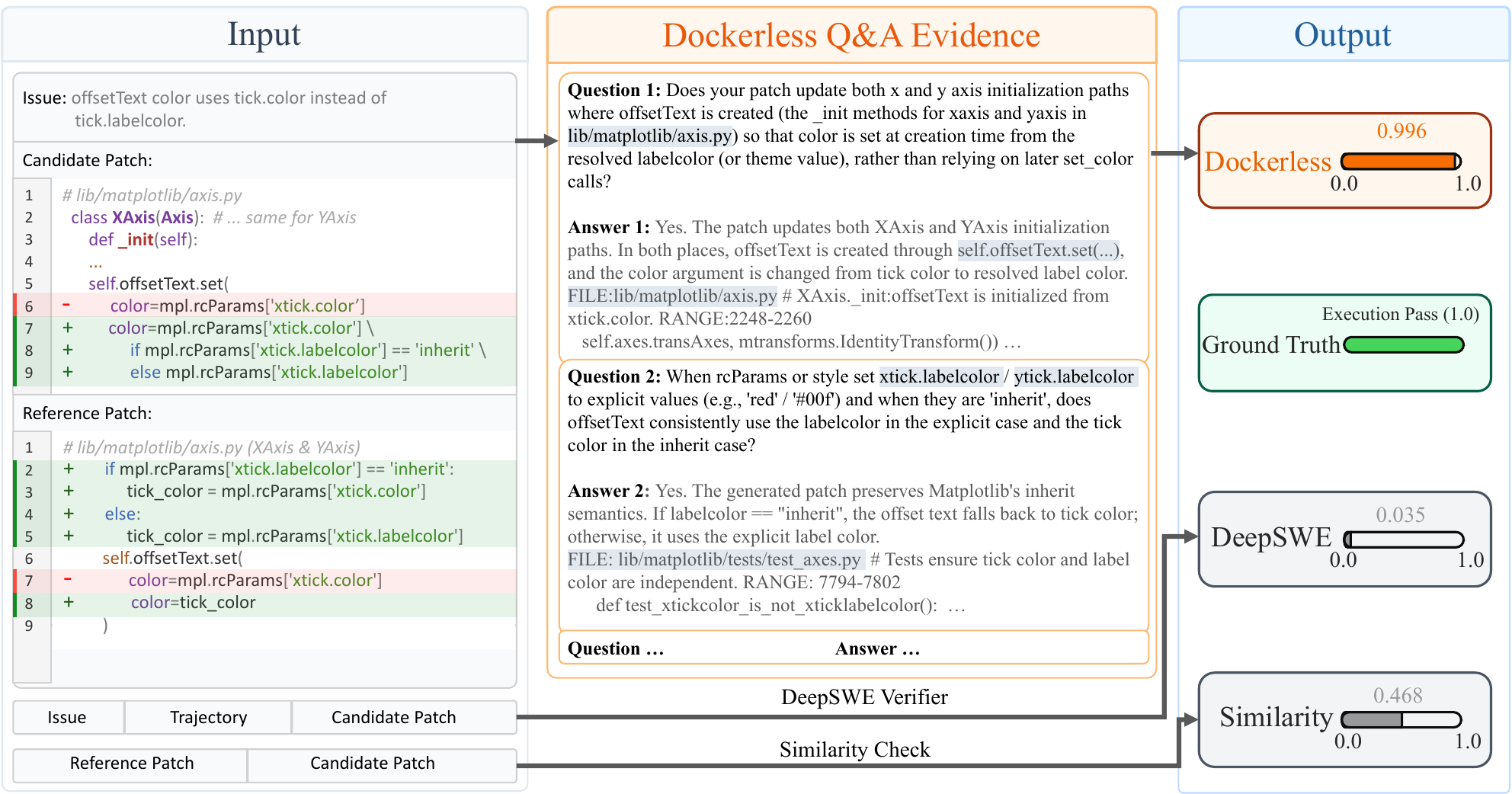}
  \caption{Representative case where the candidate patch resolves the issue but uses a different surface form from the reference patch.}
  \label{fig:case-study}
\end{figure}

\subsection{Case study}
\label{sec:case-study}

\cref{fig:case-study} shows a representative case on a \texttt{matplotlib} \texttt{offsetText} color issue: the candidate patch passes execution ($r_{\text{env}}{=}1.0$) but rewrites the fix as an inline conditional rather than the helper-variable refactor used by the reference patch.
Both baselines assign low scores: text similarity returns $0.468$, and the DeepSWE Verifier assigns $0.035$.
\approach instead dispatches one sub-agent per verification question. The gathered evidence confirms that the fix is applied to both \texttt{XAxis} and \texttt{YAxis} initialization paths in \texttt{lib/matplotlib/axis.py}, and that the \texttt{inherit} vs.\ explicit \texttt{labelcolor} semantics are preserved.
With this repository-grounded evidence, \approach scores the patch $0.996$, in agreement with the execution result. 
The case illustrates how Q\&A evidence can support a correct judgment even when the candidate patch differs substantially from the reference patch in surface form.

\section{Related Work}
\label{sec:related}

\subsection{Software Engineering Agents}
Large language models (LLMs) have rapidly evolved from generating simple code snippets~\cite{PengGGHL24,abs-2501-01329,GaoWGWZL23,shi2024code} to real-world software engineering tasks~\cite{jimenez2024swe,yang2024swe,li2025swe,chen2025swe}. SWE agents are typically post-trained with a two-stage SFT-then-RL recipe on scaffolds such as SWE-agent~\citep{yang2024swe} and OpenHands~\citep{wang2025openhands}, with SFT on curated or execution-filtered trajectories~\citep{pan2025training, jain2025r2e, yang2025swe, badertdinov2026swe, yang2025kimi} and RL driven by test-execution rewards~\citep{wei2025swerl, luo2025deepswe, golubev2025training, shao2024deepseekmath, yu2025dapo}.
A complementary line builds env-free rollout pipelines that share a single base image across repositories~\citep{sun2026swe,xu2025scalable,ludwig2026swe}, but they still constrain the agent during rollout, by exposing only a small set of static tools~\citep{xu2025scalable}, by simulating tool returns with a learned transition model~\citep{sun2026swe}, or by prompt-level restrictions on what may be executed~\citep{ludwig2026swe}.
\approach instead lets the agent issue any shell command in a minimal Linux image and receive real tool feedback, and replaces both the RFT-stage filter and the RL reward source with a single env-free agentic verifier.

\subsection{Verifiers for SWE agents}
A line of work trains LM verifiers that score a patch from a fixed prompt, ranging from execution-trained classifiers~\citep{pan2025training, jain2025r2e} and a scaled 30B mixture-of-experts critic~\citep{shum2025swe} to group-wise textual reasoning over candidates~\citep{xu2025scalable} and rubric-supervised or RL-distilled variants~\citep{wang2026rubric, luo2025deepswe}.
None of these call tools or inspect the repository at scoring time.
A more recent line frames the verifier itself as an agent, but places that agent outside the SWE patch outcome setting.
Some target domains far from SWE patches, namely mathematical reasoning~\citep{zhang2026agentv,zeng2026glimprouter} and competitive programming~\citep{ma2026scaling,hu2026line,Gao0GL25,abs-2510-17130}.
Others place the agent at rubric authoring rather than at scoring~\citep{raghavendra2026agentic}, or score intermediate trajectory steps under a fixed rubric rather than the final patch~\citep{han2026swe}.
\approach instead places the agent at SWE patch outcome scoring itself, actively exploring the repository through real tool calls before issuing a verdict.

\section{Conclusion}

In this work, we propose \approach, an agentic verifier that scores patches by actively exploring the repository, requiring no per-repository environment.
We show that \approach can serve as both the trajectory filter for SFT and the reward signal for RL, yielding a fully environment-free post-training pipeline for coding agents.
\approach outperforms prior open-source verifiers, and the resulting model matches the performance of its environment-based counterpart while requiring zero per-repository setup.
We believe that agentic, evidence-grounded verification provides a new perspective on reward modeling for code, and opens a scalable path toward post-training on the long tail of real-world repositories without reproducible execution environments.


\bibliographystyle{acl_natbib}
\bibliography{custom}

\clearpage
\appendix
\crefalias{section}{appendix}
\crefalias{subsection}{appendix}
\crefname{appendix}{Appendix}{Appendices}
\Crefname{appendix}{Appendix}{Appendices}

\section{Frontier-model env-base vs.\ env-free}
\label{app:motivation}
\label{app:motivation-eb-ef}

\paragraph{Setting.}
We motivate \approach by first checking whether env-free agent rollouts are useful at all on SWE-bench.
Four frontier models (DeepSeek-V3.2, Kimi-K2.5, GLM-5, GPT-5.4) are run under the OpenHands scaffold on SWE-bench Verified, Multilingual, and Pro.
The env-based setting uses the per-repository Docker image with the held-out test suite, exactly as in \cref{tab:main}; the env-free setting replaces it with the minimal Ubuntu image from \cref{sec:setup}, with no per-repository dependencies and test runner.
We compare resolve rate (Pass@1) under the two settings.

\begin{figure}[htbp]
  \centering
  \includegraphics[width=0.60\linewidth]{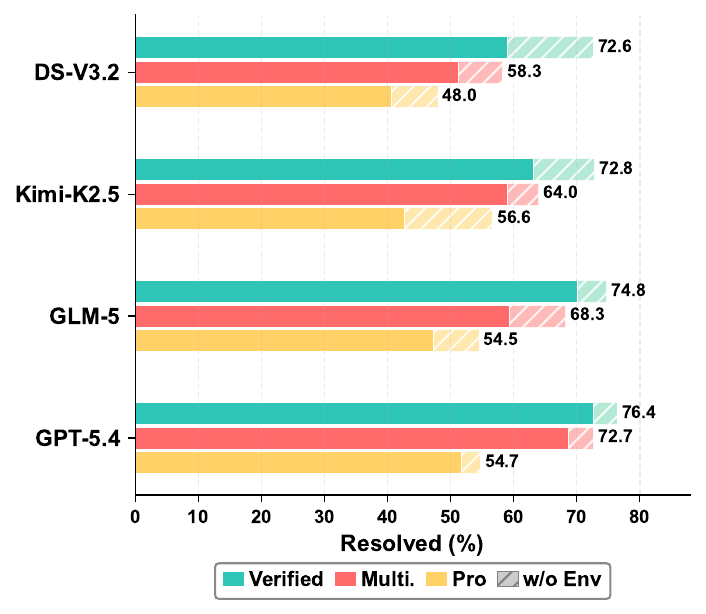}
  \caption{Frontier-model resolve rate (\%) on SWE-bench Verified, Multilingual, and Pro under env-based and env-free settings. Solid bars are env-free; hatched extensions show the additional gain from per-repository environments, so the full bar height equals the env-based score.}
  \label{fig:motivation-eb-ef}
\end{figure}

\paragraph{Removing the environment costs only a few points.}
The hatched portion of every bar in \cref{fig:motivation-eb-ef} is small: across the four models and three benchmarks, env-free evaluation costs at most $13.9$ points and on average $7.1$ points of resolve rate, with the strongest model (GPT-5.4) staying within $3.0$--$4.0$ points across all three benchmarks.
In the env-free setting, the model already solves a large fraction of issues that its env-based counterpart can solve on the same benchmarks.


\paragraph{Implications for the verifier.}
The agent side of the env-free pipeline is therefore already largely feasible: rollouts can be collected at scale on benchmarks where no per-repository Docker is available, with only a moderate quality hit.
The blocker is the verifier side, since without test execution there is no built-in correctness signal to filter rollouts or reward RL.
This is the gap \approach closes; the corresponding env-free results on our own models are reported in \cref{app:wo-env-eval}.

\section{Env-free evaluation results}
\label{app:wo-env-eval}

\paragraph{Setting.}
\cref{tab:main} reports env-based evaluation only.
For completeness, we re-evaluate the same models from \cref{tab:main} under env-free evaluation, where the agent runs in a minimal Ubuntu 22.04 LTS image with only the repository checkout at the base commit, with no per-repository Docker, no test runner, and no pre-installed dependencies (definition in \cref{sec:setup}).
Numbers are reported in \cref{tab:main-wo-env}.
\begin{table}[t]
\centering
\small
\begin{tabular}{lccc}
\toprule
\textbf{Model}
& \textbf{Verified}
& \textbf{Multilingual}
& \textbf{Pro} \\
\midrule
\texttt{SWE-Gym-7B}                & 9.0  & 5.7  & 2.2  \\
\texttt{SWE-Dev-7B}                & 8.4  & 7.0  & 2.2  \\
\texttt{SWE-Lego-8B}               & 32.0 & 17.7 & 12.7 \\
\texttt{Qwen3.5-9B}                & 50.2 & 38.3 & 26.0 \\
\midrule
\texttt{Env-SFT-9B}                & 50.0 & 36.7 & 27.2 \\
\rowcolor{black!4}
\textbf{\texttt{\approach-SFT-9B}} & \textbf{52.0} & \textbf{41.0} & \textbf{29.4} \\
\rowcolor{black!4}
\textbf{\texttt{\approach-RL-9B}}  & \textbf{53.8} & \textbf{42.3} & \textbf{30.6} \\
\bottomrule
\end{tabular}%
\caption{
Resolve rate (\%) on SWE-bench Verified, Multilingual, and Pro under env-free evaluation: the agent runs in a minimal Ubuntu image with only the repository checkout, with no per-repository Docker image and no pre-installed dependencies.
These are the same models as \cref{tab:main}, evaluated under the stricter env-free setting.
}
\label{tab:main-wo-env}
\end{table}

\paragraph{Ranking is preserved across environments.}
\approach-RL-9B remains the strongest sub-$10$B model on every benchmark under env-free evaluation ($53.8$, $42.3$, $30.6$), ahead of \approach-SFT-9B by $1.2$--$1.8$ points and of the env-based-SFT baseline \texttt{Env-SFT-9B} by $3.8$, $5.6$, and $3.4$ points on Verified, Multilingual, and Pro respectively.
The aggregate story from \cref{tab:main} therefore carries over to env-free deployment: a fully env-free pipeline (\approach-SFT-9B and \approach-RL-9B trained without test execution) still produces the strongest model when the deployment setting also forbids test execution.

\paragraph{\approach-trained models are more robust to env-free deployment.}
Comparing each model's env-based score in \cref{tab:main} to its env-free score in \cref{tab:main-wo-env}, the average drop is $9.4$ points for \texttt{Env-SFT-9B}, $7.1$ points for \approach-SFT-9B, and $6.8$ points for \approach-RL-9B.
The env-free-trained models thus suffer a smaller env-base-to-env-free gap than the env-based-trained baseline, which is the expected direction: a model that has been trained on env-free rollouts has seen the same distribution it is evaluated on at deployment, while the env-based-trained baseline is exposed to a distribution shift.
The same trend holds for the open-source SFT specialists, whose absolute scores are too low to draw strong conclusions but whose gaps lie in the same range.

\section{Dataset Construction}
\label{app:data}

\subsection{Agentic verifier training data}
\label{app:data-verifier}

We construct a training corpus from execution-labeled patches in SWE-Gym~\citep{pan2025training} and Multi-SWE-RL~\citep{zan2026multi}, with $r^\star \in \{0, 1\}$ being the verdict obtained from running the held-out unit tests on the candidate patch.
These datasets are disjoint from our verifier evaluation benchmark built from SWE-bench Verified and Multi-SWE-bench Flash.
For each source example, a strong frontier teacher model (GLM-5) proposes one or more candidate $(Q\text{+}A\text{+Judge trajectory}, \hat{r})$ tuples via the same workflow used at inference (\cref{sec:agentic-verifier}).
We then keep only tuples whose predicted verdict $\hat{r}$ matches $r^\star$, and apply two additional cleaning passes: we discard answer trajectories with fewer than $4$ or more than $30$ turns, and we remove malformed or interrupted exchanges.
Finally, we cap the negative-to-positive sample ratio at $4{:}1$ to mitigate class imbalance, following the recipe of \citet{shum2025swe}.
The resulting corpus covers $3.7$K unique issues.
Each training example bundles one question-generation trajectory, $K$ sub-agent answer trajectories, and one final-judgment trajectory, all generated by the same teacher.
We do not enforce a target ratio across these three sub-tasks; they are jointly trained on the same backbone under \cref{eq:rm-loss}.
Both inputs (the candidate patch text and the rendered Q+A context) are truncated to $10{,}000$ characters before being fed to the model.

\subsection{Env-free rollout data}
\label{app:data-rollout}

The pool of env-free rollouts used to construct $\mathcal{D}_{\text{RFT}}$ is collected on SWE-Rebench-v2~\citep{badertdinov2026swev2}.
Starting from the OpenHands agent and a minimal Linux image, the agent receives only the issue $x$ and the repository at the base commit, with no per-repository Docker image; OpenHands tools remain available and may include execution feedback from running standard developer utilities.
We collect a pool of $16$K rollouts at sampling temperature $1.0$, from which the downstream filter selects $4$K trajectories globally ranked by \approach.

\subsection{Verifier evaluation benchmark}
\label{app:rm-bench}

We construct a balanced trajectory-level verifier benchmark to evaluate \approach against prior verifiers (\cref{tab:rm-bench}).
The benchmark contains $500$ samples drawn from SWE-bench Verified and $276$ samples drawn from Multi-SWE-bench Flash, with positive and negative labels balanced within each split.
Trajectories are collected from several models running under the SWE-agent and OpenHands scaffolds in a $1{:}1$ split, and each (issue, candidate patch) pair is labeled positive or negative via standard evaluation inside the per-repository Docker environment with held-out tests; positive and negative labels are balanced $1{:}1$ within each split.

\section{Training Details}
\label{app:training}


\subsection{Agentic verifier}
\label{app:training-verifier}

The agentic verifier is initialized from Qwen3.5-9B~\citep{team2026qwen35} and fine-tuned with standard next-token cross-entropy on the filtered trajectories described in \cref{app:data-verifier} (\cref{eq:rm-loss}).
We report results from the best checkpoint, reached at $150$ optimizer steps on a held-out validation split.
We use AdamW with learning rate $1.0\mathrm{e}{-5}$ (cosine decay to $1.0\mathrm{e}{-6}$, warmup ratio $0.05$), weight decay $0.01$, batch size $256$, and maximum sequence length $32{,}768$.

At inference, we serve the trained model via vLLM with the OpenAI-compatible API.
The verifier generates $2$--$4$ verification questions per scoring call, with one sub-agent dispatched per question to explore the repository in parallel.
At the answer position, we read the logits of the ``$0$'' and ``$1$'' verdict tokens and convert them into the dense score via softmax (\cref{sec:agentic-verifier}).

\subsection{Env-free SFT}
\label{app:training-sft}

Each candidate rollout's final patch is scored by \approach with $M{=}2$ independent agentic passes; we report the mean dense score and discard any pass that fails (e.g., due to inference-time errors or timeouts).
We then build $\mathcal{D}_{\text{RFT}}$ by selecting the top-ranked $4$K rollouts globally from the $16$K pool.
The SFT model is initialized from Qwen3.5-9B~\citep{team2026qwen35} and trained with standard maximum-likelihood on $\mathcal{D}_{\text{RFT}}$.
We use the same AdamW configuration as the verifier (\cref{app:training-verifier}), trained for $3$ epochs.

\subsection{Env-free RL}
\label{app:training-rl}

We initialize the RL policy from the SFT model produced in \cref{app:training-sft} and run GRPO~\citep{shao2024deepseekmath} with \approach as the per-rollout reward source.
For each issue $x$, we sample a group of $G{=}8$ rollouts and score every rollout with $M{=}2$ independent agentic passes through \approach; failed passes are dropped and the remaining dense scores are averaged to form $r_\phi(x, y_i)$.
The group-normalized advantages from \cref{eq:grpo-advantage} drive the policy update following the standard GRPO objective; no test execution is performed at any step.
We use actor learning rate $2.0\mathrm{e}{-6}$, training batch size $64$, PPO mini-batch size $64$, $8$ responses per prompt, clipping range $[0.2, 0.27]$, entropy coefficient $0$, KL coefficient $0$ (no KL loss), maximum $150$ turns per rollout, and sampling temperature $1.0$.
We train for $50$ RL steps in total.

\section{Per-Language Analysis}
\label{app:per-language}

\begin{figure}[t]
  \centering
  \includegraphics[width=0.60\linewidth]{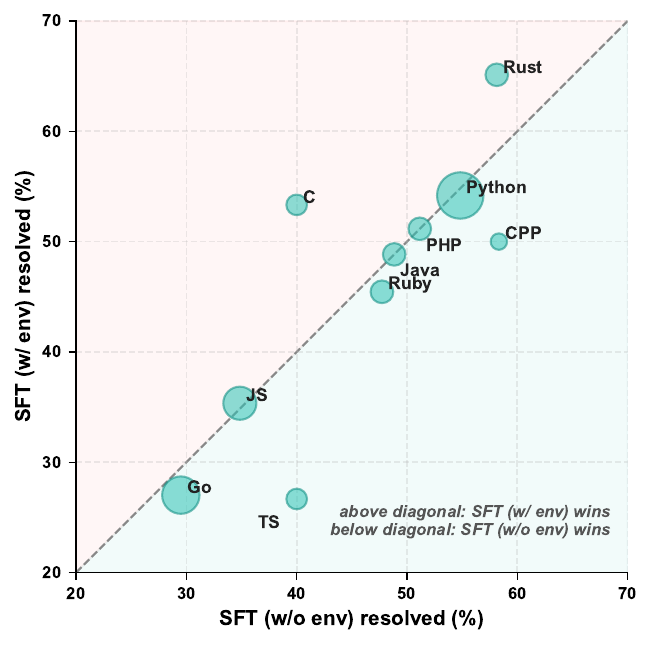}
  \caption{Per-language comparison of SFT (w/o env) and SFT (w/ env) resolve rate, aggregated across SWE-bench Verified, Multilingual, and Pro. Bubble size encodes the number of test instances per language. Points above the diagonal mark languages where SFT (w/ env) wins, below the diagonal where SFT (w/o env) wins.}
  \label{fig:per-lang-scatter}
\end{figure}

While the aggregate SFT (w/o env) vs.\ SFT (w/ env) gap in \cref{tab:main} is small, the per-language breakdown (\cref{fig:per-lang-scatter}) is uneven, and the unevenness has a consistent pattern.

The two languages where SFT (w/ env) clearly wins are also the two compilation-heavy ones in the benchmark: Rust ($+7.0$) and C ($+13.3$).
On the remaining high-volume languages (Python, Go, JavaScript, Java, PHP), the two settings stay within $\pm 2.5$ points of each other.
The two large gaps below the diagonal (TypeScript $-13.3$, C++ $-8.3$) come from splits with only $30$ and $12$ instances, so we do not read them as evidence either way.

We attribute the Rust/C gap to compiler diagnostics being available only inside the per-repository environment: env-base trajectories can observe type errors and link failures at intermediate steps, while env-free trajectories must infer the same information from the source alone.
This is consistent with the broader claim that the residual value of env-base supervision is concentrated in compiler signal, not in test execution per se, although Rust and C are only two languages and we do not treat this as proof.

The takeaway for \cref{tab:main} is that the headline ``env-free matches env-base'' holds across the high-volume languages that dominate the aggregate, but understates a real $7$--$13$ point gap on compilation-heavy languages.
Closing that gap likely requires surfacing compiler-style feedback inside the env-free pipeline rather than scaling env-base data further, which we leave to future work.

\section{Latency distribution}
\label{app:latency}

\begin{figure}[t]
  \centering
  \includegraphics[width=0.70\linewidth]{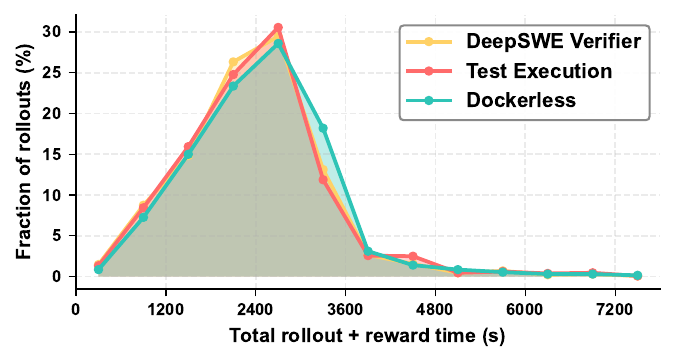}
  \caption{Distribution of total per-rollout wall-clock time (rollout + reward) under three reward sources, on $7680$ rollouts collected during RL training. All three sources produce near-identical distributions; the long right tail is set by slow rollouts, not by the verifier.}
  \label{fig:latency-distribution}
\end{figure}

\Cref{fig:latency-distribution} complements the mean numbers in \cref{sec:latency-analysis} by showing the full distribution of per-rollout wall-clock time (rollout + reward) for the three reward sources.
The three distributions overlap almost completely: a single mode around $2400$--$3000$s and a heavy tail extending to the hard timeout.
The choice of reward source shifts the mean by less than $150$s, well inside the spread of the rollout distribution itself, so the end-to-end RL step is bottlenecked by the slowest rollouts in each group rather than by reward latency.

\section{Prompt Templates}
\label{app:prompts}

We list below the full prompts used by \approach.

\subsection{Question generation prompt}
\label{app:prompt-question}

The question generator takes the issue description and the reference patch and emits $2$--$4$ diagnostic questions, each tagged with one of four categories (location, behavior, test evidence, edge case) and accompanied by a short rationale.

\begin{promptbox}{Stage 1: Question Generation Prompt}
  \#\# Role\\
  You are an expert code reviewer specialized in analyzing patches across different programming languages.

  \#\# Task\\
  Generate \textbf{2--4 focused questions} that would provide strong evidence for judging whether a candidate's patch correctly fixes the issue. Your questions should help distinguish a correct fix from an incorrect or incomplete one.

  \#\# Input\\
  \#\#\# Issue Description\\
  \{issue\_description\}

  \#\#\# Golden Patch (Correct Solution)\\
  \{golden\_patch\}

  \#\# Analysis Framework\\
  Before generating questions, briefly analyze: (1) Bug Type, (2) Failure Scenario, (3) Success Criteria.

  \#\# Question Categories\\
  Your questions MUST fall into one of these categories:
  \begin{itemize}\setlength\itemsep{0pt}
    \item \textbf{Location}: identify WHERE the fix should be applied.
    \item \textbf{Behavior}: understand WHAT the code should do.
    \item \textbf{Test Evidence}: find HOW to verify correctness.
    \item \textbf{Edge Cases}: discover WHAT ELSE might break.
  \end{itemize}

  \#\# Output Format\\
  Output a single JSON object with an \texttt{analysis} field (bug\_type, failure\_scenario, success\_criteria) and a \texttt{questions} list, each item carrying \texttt{id}, \texttt{category}, \texttt{question}, and \texttt{rationale}.

  \#\# Key Principle\\
  Focus on evidence that can distinguish between a complete fix, a partial fix, and an incorrect fix.
\end{promptbox}

\subsection{Sub-agent exploration prompt}
\label{app:prompt-explore}

The sub-agent runs a ReAct-style read-only shell loop. It is configured with a fixed system message and a per-question instance template, and emits a final answer through a \texttt{SUBMIT\_ANSWER} heredoc.

\begin{promptbox}{Stage 1: Sub-agent System Message}
  You are a code exploration assistant that answers questions by running shell commands. You analyze codebases in various programming languages (Python, Java, JavaScript, Go, C++, etc.).

  \#\# Response Format\\
  Every response MUST contain: (1) a \textbf{THOUGHT} section explaining your reasoning, and (2) exactly \textbf{ONE} bash command in a code block.

  \#\# Rules\\
  This is a READ-ONLY task; do NOT modify any files. Execute ONE command per response. Use non-interactive commands only. Chain commands with \texttt{\&\&} or \texttt{||} when needed in a single block.
\end{promptbox}

\begin{promptbox}{Stage 1: Sub-agent Per-Question Instance Template}
  \textless question\textgreater\\
  Answer the following question by exploring the codebase.

  \#\#\# Issue Description\\
  \{issue\_description\}

  \#\#\# Golden Patch\\
  \{golden\_patch\}

  \#\#\# Question\\
  \{question\}\\
  \textless /question\textgreater

  \textless instructions\textgreater\\
  \#\# Goal\\
  Find concrete evidence that helps verify if a patch correctly fixes the issue. This is a READ-ONLY task.

  \#\# Search Strategy (priority order)\\
  (1) Test Evidence: find tests that validate the correct behavior. (2) Error Location: find where the error or failure occurs. (3) Code Context: understand the code being patched, including definitions and call sites. (4) Documentation \& Configuration: find expected-behavior documentation.

  \#\# Final Answer Format\\
  When evidence is sufficient, submit the answer via a \texttt{cat \textless \textless 'ANSWER\_EOF'} block containing: a \texttt{SUBMIT\_ANSWER} marker, a Direct Answer, Key Findings, and one or more \texttt{FILE} / \texttt{RANGE} code-evidence blocks.

  \#\# Important Notes\\
  Focus on concrete evidence (actual code, tests, assertions). Prioritize test cases. Include file paths and line numbers for all evidence. Be language-aware but use language-agnostic search patterns.\\
  \textless /instructions\textgreater
\end{promptbox}

\subsection{Final scoring prompt}
\label{app:prompt-score}

The judge model conditions on the issue, the reference patch, the candidate patch, and the Q\&A context collected by the sub-agents, and produces a single binary verdict token whose logits define the continuous score $r_\phi(x, y)$.

\begin{promptbox}{Stage 2: Final Scoring Prompt}
  \#\# Role:\\
  You are an expert code reviewer.

  \#\# Task:\\
  Evaluate whether the ``Generated Patch'' solves the issue described in the ``Issue Description'' correctly, using the Q\&A context gathered in Stage~1 and the golden patch (a correct solution) as a reference.

  \#\# Issue Description:\\
  \{issue\_description\}

  \{context\}

  \#\# Golden Patch:\\
  \{golden\_patch\}

  \#\# Generated Patch:\\
  \{generated\_patch\}

  \#\# Response Format:\\
  You must follow the response format strictly:

  \textless answer\textgreater[0 for not solving the issue, 1 for solving the issue]\textless /answer\textgreater
\end{promptbox}

\subsection{Zero-shot LLM-as-judge prompt}
\label{app:prompt-judge}

For the frontier LLM-as-judge baselines in \cref{tab:rm-bench} (DeepSeek-V3.2, Kimi-K2.5, GLM-5, GPT-5.4), we query each model zero-shot with the issue description, the reference (golden) patch, and the candidate patch, and parse the binary verdict from the \texttt{<answer>} tag.

\begin{promptbox}{Zero-shot LLM-as-Judge Prompt}
  \#\# Role:\\
  You are an expert code reviewer.

  \#\# Task:\\
  Your task is to evaluate whether the ``Generated Patch'' solves the issue described in the ``Issue Description'' correctly based on the issue description and a golden patch (a reference correct solution).

  \#\# Issue Description:\\
  \{issue\_description\}

  \#\# Golden Patch:\\
  \{golden\_patch\}

  \#\# Generated Patch:\\
  \{generated\_patch\}

  \#\# Response Format:\\
  You must follow the response format strictly:

  \textless answer\textgreater[0 for not solving the issue, 1 for solving the issue]\textless /answer\textgreater
\end{promptbox}
\end{document}